\documentclass[12pt]{article}
\usepackage{amssymb}

\usepackage{amsmath}

\newtheorem{theorem}{Theorem}
\newenvironment{proof}[1][Proof]{\textbf{#1.} }{\ \rule{0.5em}{0.5em}}

\input{tcilatex}

\newcommand{\benumerate}{\begin{enumerate}}
\newcommand{\eenumerate}{\end{enumerate}}

\newcommand{\bitemize}{\begin{itemize}}

\newcommand{\eitemize}{\end{itemize}}
\newcommand{\der}[2]{\frac{\partial #1}{\partial #2}}
\newcommand{\derbar}[2]{\frac{\partial #1}{\partial \bar{#2}}}
\newcommand{\dersec}[2]{\frac{\partial^{2} #1}{\partial #2^{2}}}
\newcommand{\derthree}[2]{\frac{\partial^{3} #1}{\partial #2^{3}}}
\newcommand{\derfour}[2]{\frac{\partial^{4} #1}{\partial #2^{4}}}
\newcommand{\derfive}[2]{\frac{\partial^{5} #1}{\partial #2^{5}}}

\newcommand{\dermixd}[3]{\frac{\partial^{2} #1}{\partial #2 ~\partial #3}}
\newcommand{\binomial}[2]{\left( \parbox[c]{0.2cm}{#1 \\ #2}  \right)}

\newcommand{\dertot}[2]{\frac{d #1}{d #2}}
\newcommand{\dertotsec}[2]{\frac{d^{2} #1}{d #2^{2}}}
\newcommand{\poisson}[4]{\left \{#1,#2 \right \}_{#3,#4}}

\newcommand{\paperO}[6]{#1, #2, {\em ``#3''}, #4, {\bf #5},  (#6)}

\newcommand{\book}[5]{#1, #2,{\em ``#3''}, (#4, #5)}
\newcommand{\paperOJPA}[5]{#1, #2, #3, {\bf #4}, #5}
\newcommand{\paperOJPAr}[4]{#1, #2, {\bf #3}, #4}

\begin{document}


\title{Quasi-classical $\bar{\partial}$-dressing approach to the
weakly dispersive KP hierarchy}

\author{Boris Konopelchenko\footnote{Supported in part by the COFIN
PRIN ``SINTESI'' 2002.}~\footnote{e-mail konopel@le.infn.it}~ and
Antonio Moro$^{\ast}$\footnote{e-mail antonio.moro@le.infn.it} \\
{\em Dipartimento di Fisica dell'Universit\`{a} di Lecce} \\
{\em and INFN, Sezione di Lecce, I-73100 Lecce, Italy}}
\maketitle

\begin{abstract}
Recently proposed quasi-classical
$\bar{\partial}$-dressing method provides a systematic approach to
study the weakly dispersive limit of integrable systems. We apply the
quasi-classical $\bar{\partial}$-dressing method to describe
dispersive corrections of any order. We show how to calculate
the $\bar{\partial}$ problems at any order for a rather general class
of integrable systems, presenting explicit results for the KP hierarchy
case. We demonstrate the stability of the method at each order. We construct
an infinite set of commuting flows at first order which allow a
description analogous to the zero order (purely dispersionless) case,
highlighting a Whitham type structure. Obstacles for the construction of the higher order
dispersive corrections are also discussed.

\end{abstract}

\section{Introduction}
\label{secintro}
During the last decade a great interest has been focused on the 
dispersionless limit of integrable nonlinear dispersive
systems. They arise in various contexts of physics and mathematical
physics: topological field theory and
strings, matrix models, interface 
dynamics, nonlinear optics, conformal mappings
theory\cite{witten1}-\cite{boyarsky}.
Several methods and approaches have been used in order to study the
features of this type of systems, like the
quasi-classical Lax pair representation with its close relationship with
the Whitham universal hierarchy\cite{krich1}-\cite{taktak2};
the hodograph transformations to calculate particular and interesting
classes of exact solutions (rarefaction and shock waves)\cite{kodama-exact}; the
quasi-classical version of the inverse scattering method allowing to
analyze various 1+1-dimensional systems\cite{1+1dim}. Moreover,
the recent formulation of the quasi-classical 
$\bar{\partial}$-dressing method\cite{kon2}-\cite{ragnisco}
introduces a more general and 
systematic approach to multidimensional integrable systems with weak
dispersion, which preserves the power of the standard method. It allows to
build integrable systems and solutions simultaneously; that is a non
trivial fact, because of the difficulty to obtain
exact explicit solutions\cite{lorenzoni}. In addition, new
interesting interrelations have been revealed, like an intriguing connection
with the quasi-conformal mapping theory, a strong similarity with the
theory of semi-classical approximation to quantum
mechanics\cite{maslov} and geometric asymptotics methods to calculate
wave corrections to geometrical optics\cite{stern}.
While the method has been rather widely discussed for purely
dispersionless limit of some famous integrable 
hierarchies (KP, mKP, 2DTL)\cite{kon2,kon3,ragnisco,kon3bis,kon4} the study of the dispersive
corrections is open yet. The latter is an interesting and
challenging question in the context where one would like to be able to
investigate the properties of the full dispersive system through an
approximation theory.

A main goal of the present paper is to extend the quasi-classical
$\bar{\partial}$-dressing method to higher order dispersive
corrections, starting an investigation of the general properties at
the dispersive orders of an integrable hierarchy.  We will show that the quasi-classical
$\bar{\partial}$-dressing method works as effectively as in the pure
dispersionless case, and the first order admits a
construction parallel to leading order case. We present all calculations for
the weakly dispersive KP hierarchy as illustrative example.
In this paper we do not consider the problem of construction of bounded
solutions. So, we do not care about secular terms which usually appear
in asymptotic expansion\cite{whithambook}. 

We briefly present the standard $\bar{\partial}$-method in the
Sec.\ref{secstand} and the quasi-classical
$\bar{\partial}$-method in the Sec.\ref{secdispless}. In the
Sec.\ref{secanyord}, we calculate the quasi-classical
$\bar{\partial}$-problems at any order in the dispersive parameter, giving the explicit formulas at fourth order. In the
Sec.\ref{secfirst} we recall the method at the first order\cite{kon2}, and
prove a theorem allowing us to build an infinite set of flows
reproducing the first corrections to the dKP equation.
In the Sec.\ref{sechigher} we present the generalization of the
$\bar{\partial}$-dressing method for higher orders. The
Sec.\ref{secLax} 
contains some considerations about the possibility to describe the
first order introducing a certain generalization of the Whitham hierarchy . 
Finally, we present some concluding remarks.

\section{The standard $\bar{\partial}$-dressing method.}
\label{secstand}
The standard $\bar{\partial}$-dressing method is a powerful
procedure allowing to construct multidimensional integrable hierarchies
of equations and their solutions. In this section we outline the main ideas of
the method\cite{manakov,kon1}. The $\bar{\partial}$-dressing method
is based on the nonlocal $\bar{\partial}$-problem

\begin{equation}
\label{DBAR non-local}
\derbar{\chi(z,\bar{z};t)}{z}= \int_{\mathbb{C}} d\mu
\wedge d\bar{\mu}\: \chi(\mu,\bar{\mu};t)\: g(\mu,t)\:
 R_{0}(\mu,\bar{\mu};z,\bar{z}) \:g^{-1}(z,t) ,
\end{equation}

where $z$ is the spectral parameter and $\bar{z}$ its complex
conjugate, $t = (t_{1},t_{2},t_{3},\dots)$  the vector of the times,
$\chi(z,\bar{z};t)$ a complex-valued function on
the complex plane $\mathbb{C}$ and $R_{0}(\mu,\bar{\mu};z,\bar{z})$ is the
$\bar{\partial}$-data.

The $\bar{\partial}$-equation~(\ref{DBAR non-local}) encodes all
informations about integrable hierarchies. It is assumed  that the
problem~(\ref{DBAR non-local}) is uniquely solvable. Usually, one
consider its solutions with canonical
normalization, i.e.

\begin{equation}
\label{boundary}
\chi \rightarrow  1 + \frac{\chi_{1}}{z} +
\frac{\chi_{2}}{z^{2}}+ \dots ,
\end{equation}

as $z \rightarrow \infty$.

The particular form of the function $g(z,t)$ sets an integrable
hierarchy. Specifically, in order to describe the KP hierarchy one has
to take 

\begin{equation}
g(z,t) = e^{S_{0}(z,t)}, \;\;\;S_{0}(z,t) = \sum_{k=1}^{\infty} z^{k} t_{k}.
\end{equation}

Following a standard procedure\cite{kon1}, it is possible to construct
an infinite 
set of linear operators $M_{i}$, providing an
infinite set of linear equations

\begin{equation}
\label{M-operators}
M_{i} \chi = 0,
\end{equation}

which are automatically compatible. Equations~(\ref{M-operators}) provide
us with the corresponding integrable hierarchy.
For example, the pair of operators (Lax pair)

\begin{eqnarray}
M_{2} &=& \der{}{y} - \dersec{}{x} - u(x,y,t), \nonumber \\
M_{3} &=& \der{}{t} - \derthree{}{x} - \frac3 2 u
\der{}{x} - \frac3 2 u_{x} - \frac3 4 \partial^{-1}_{x} u_{y}, \nonumber
\end{eqnarray}

with the notations

\begin{eqnarray}
t_{1} = x ; \;\;\;  t_{2} = y ;\;\;\; t_{3} = t; \nonumber \\
u_{t_{i}} = \der{u}{t_{i}}; \;\;\; \partial^{-1}_{x} = \int_{-\infty}^{x} dx,\nonumber
\end{eqnarray}

provides us with the KP
equation (sometimes called ``{\em dispersionfull}'' KP)

\begin{equation}
\label{dispfull}
\dermixd{u}{t}{x} = \der{}{x}\left(\frac1 4 \derthree{u}{x} + \frac3 2
u \der{u}{x}\right) + \frac3 4 \dersec{u}{y}.
\end{equation}

$\bar{\partial}$-dressing method allow us to construct a wide class of
exact solutions of the KP equation and the whole KP hierarchy.

\section{Quasi-classical $\bar{\partial}$-dressing method.}
\label{secdispless}
\subsection{Dispersionless limit.}
In order to illustrate the quasi-classical
$\bar{\partial}$ formalism, we, following Ref.\cite{kon2}, introduce
slow variables 
$t_{i} = \frac{T_{i}}{\epsilon}$, $\partial_{t_{i}}=\epsilon
\partial_{T_{i}}$ and assume that $g(z,T;\epsilon) =
e^{\frac{S_{0}(z,T)}{\epsilon}}$, where $\epsilon$ is a small
dispersive parameter. We are looking for solutions of the
$\bar{\partial}$-problem in the form 

\begin{equation}
\label{asymptotic0}
\chi\left(z,\bar{z};\frac{T}{\epsilon}\right) =
\tilde{\chi}(z,\bar{z};T;\epsilon) \exp\left(\frac{\tilde{S}(z,\bar{z};T)}{\epsilon}\right),
\end{equation}

where $\tilde{\chi}(z,\bar{z};T;\epsilon) =
\sum_{n=0}^{\infty}\varphi_{n}(z,\bar{z};T) \epsilon^{n}$ is an
asymptotic expansion. The function
$\tilde{S}$ has the following expansion

\begin{equation}
\tilde{S} \rightarrow \frac{\tilde{S}_{1}}{z} +
\frac{\tilde{S}_{2}}{z^{2}} + \frac{\tilde{S}_{3}}{z^{3}}+\dots, ~~z
\rightarrow \infty.
\end{equation}

Let us consider the $\bar{\partial}$-kernel of the following, quite
general form:

\begin{equation}
\label{kernel1ss}   
R_{0}(\mu,\bar{\mu};z,\bar{z}) = \frac{i}{2} \sum_{k=0}^{\infty} (-1)^{k}\:
\epsilon^{k-1} \: \delta^{(k)}(\mu-z) \Gamma_{k}(z,\bar{z}).
\end{equation}

The derivative $\delta$-functions $\delta^{(k,n)}$ are defined, in the
standard way,
as the distributions such that

\begin{equation}
- \frac{1}{2i}\int_{\mathbb{C}} d\mu \wedge d\bar{\mu}\:\delta^{(k,n)}(\mu-z)
f(\mu,\bar{\mu}) = (-1)^{k+n}\: \frac{\partial^{k+n}
f(z,\bar{z})}{\partial z^{k} \partial \bar{z}^{n}},
\end{equation}

and $\delta^{(k)} := \delta^{(k,0)}$.

Introducing

\begin{equation}
S(z,\bar{z};T) =
S_{0}(z;T)+\tilde{S}(z,\bar{z};T),
\end{equation}

one has, in the KP case,

\begin{equation}
\label{Sexpansion}
S = \sum_{k=1}^{\infty} z^{k} T_{k} + \sum_{j=1}^{\infty}
\frac{\tilde{S}_{j}}{z^{j}}, ~~ z \rightarrow \infty.
\end{equation}

A straightforward
calculation for the problem~(\ref{DBAR non-local}) with the kernel~(\ref{kernel1ss}) gives the
dispersionless (classical) limit of $\bar{\partial}$-problem

\begin{equation}
\label{classical}
\derbar{S}{z} = W\left(z,\bar{z}, \der{S}{z} \right), 
\end{equation}

where

\begin{equation}
W =\sum_{k=0}^{\infty}(-1)^{k}
\left (\der{S}{z}\right)^{k} \Gamma_{k}. 
\end{equation}

In this limit the role of the nonlocal linear $\bar{\partial}$-problem~(\ref{DBAR non-local}) is held by the nonlinear
$\bar{\partial}$-equation~(\ref{classical}), that is a local 
nonlinear equation of Hamilton-Jacobi type. The latter encodes all
informations about the integrable systems, like in the dispersionfull case.
Nevertheless, there are substantial technical differences in the
dressing procedure. In particular, the quasi-classical $\bar{\partial}$-method is based crucially on the {\em Beltrami equation}'s properties. 
We present them in the next subsection.

\subsection{The Beltrami equation ($BE$).}

The {\em Beltrami equation} (BE) has the form
 
\begin{equation}
\derbar{\Psi}{z} = \Omega(z,\bar{z})
\der{\Psi}{z},
\end{equation}

where  $z \in \mathbb{C}$. Under certain conditions, it has the following
properties (see e.g.\cite{vekua}):

\begin{enumerate}
\label{belprop1}
\item If $\Omega$  satisfies $|\Omega|\:{<}\:k<1$, then the only solution
of~$BE$ such that $\derbar{\Psi}{z}$ is locally $L^{p}$ for
some $p>2$, and such that $\Psi$ vanishes at some point of the
extended plane $\mathbb{C}^{*}$ is $\Psi \equiv 0$ ({\em Vekua's theorem}).
\label{belprop2}
\item If $\Psi_{1},\Psi_{2},\dots,\Psi_{N}$ are solutions of $BE$,
a differentiable function \\ $f(\Psi_{1},\Psi_{2},\dots,\Psi_{N})$ with
arbitrary $N$ is a solution of~$BE$ too.
\label{belprop3}
\item The solutions of the $BE$ under the previous
conditions give quasi-conformal maps with the complex dilatation $\Omega(z,\bar{z})$\cite{lehto}.
\end{enumerate}
  
It is assumed, in all our further constructions, that these properties
of $BE$ are satisfied.

\subsection{Zero order.}
The zero order case has been widely discussed, for
different integrable hierarchies in various
papers\cite{kon2,kon3,ragnisco,kon3bis,kon4}. Here, we outline
the procedure for the KP hierarchy.
The first crucial observation is that the symmetries of the
$\bar{\partial}$-problem at leading 
order in $\epsilon$, for any time $T_{j}$, i.e. $\delta S =
\der{S}{T_{j}} \delta T_{j}$, are given by the $BE$

\begin{equation}
\derbar{}{z} \left(\der{S}{T_{j}}\right) = W' \der{}{z}\left(\der{S}{T_{j}}\right). 
\end{equation}

We assume that the complex dilatation $W'$ has good properties
required for the application of the Vekua's
theorem. From~(\ref{Sexpansion}) it follows that

\begin{equation}
\der{S}{T_{j}} = z^{j} + \frac{1}{z} \der{\tilde{S}_{1}}{T_{j}} +
\frac{1}{z^2} \der{\tilde{S}_{2}}{T_{j}} + \dots, ~~z \rightarrow \infty.
\end{equation}

Using the $BE$ properties, one gets the following equations in time
variables \cite{kon2,kon3}

\begin{eqnarray}
\label{Stimey}
&&\der{S}{y} - \left(\der{S}{x}\right)^{2} - u_{0}(x,y,t) = 0,\\
\label{Stimet}
&&\der{S}{t} - \left(\der{S}{x}\right)^{3} - \frac3 2 u_{0} \der{S}{x} -
V_{0}(x,y,t) = 0,
\end{eqnarray} 

where the notation

\begin{equation}
 T_{1} = x ; \;\;\;  T_{2} = y ;\;\;\; T_{3} = t, 
\end{equation}

is adopted and
\begin{equation}
u_{0} = - 2 \der{\tilde{S}_{1}}{x} ; \; \der{V_{0}}{x} = \frac3 4 \der{u_{0}}{y}.
\end{equation}

Indeed, due to the property~2 of $BE$, the left hand sides
of equations~(\ref{Stimey}) and (\ref{Stimet}) are solutions of $BE$  and since
they vanish at $z \rightarrow \infty$, then, by virtue of the {\em
Vekua's theorem}, they vanish identically on whole complex plane.  
Equations~(\ref{Stimey}) and (\ref{Stimet}) are automatically compatible and  by
expansion in $1/z$-power, according to the standard
$\bar{\partial}$-dressing procedure, one obtains the well known
dispersionless KP (dKP) equation ( or Zabolotskaya-Khokhlov equation )

\begin{equation}
\label{dKP}
\dermixd{u_{0}}{t}{x} = \frac3 2 \der{}{x}\left(u_{0} \der{u_{0}}{x} \right)
+ \frac3 4 \dersec{u_{0}}{y}.
\end{equation}


Usually the dKP equation and its dispersive corrections
are obtained, by a slow times limit, from dispersionfull
KP equation~(\ref{dispfull}), 
inserting the following asymptotic expansion in terms of a small
dispersive parameter

\begin{equation}
\label{formal-expansion}
u\left (\frac{T}{\epsilon}\right) = \sum_{n=0}^{\infty} u_{n}(T) \epsilon^{n}.
\end{equation}

\label{secKrich}
We stress that the formal expansion~(\ref{formal-expansion}) implies,
in general,
the appearance of secular terms. Usually one imposes additional
conditions on equations~(\ref{Stimey}) and (\ref{Stimet}) to ensure
the uniform validity of this expansion~\cite{whithambook}. In this
paper we do not care about boundedness of solutions
(\ref{formal-expansion}) and, consequently, about secular terms.

Actually, as it is well known, equations~(\ref{Stimey}) and (\ref{Stimet}) are the first ones of an
infinite set of equations in all time variables $T_{n}$, with $n \in \mathbb{N}\setminus \left \{ 0 \right \}$, reproducing the whole KP
hierarchy. 
Assuming

\begin{equation}
W\left(z,\bar{z},T\right) = \theta(r - |z|)
V\left(z,\bar{z}, \der{S}{z} \right),
\end{equation}

 with $r>0$, the function
$S$ is an analytic function  outside the circle
${\cal D} = \left \{z \in \mathbb{C} \;|\; |z|<r \right \}$. That is in agreement with the asymptotic behavior~(\ref{Sexpansion}). 
At last, the dKP hierarchy can be written in compact form as
follows

\begin{equation}
\label{omega1}
\der{S}{T_{n}}\left(z,T\right) = \Omega_{n}\left(p(z,T),T\right);\;n
\geq 1, 
\end{equation} 

where $z \in \mathbb{C} \setminus {\cal D}$, and
\begin{equation}
p := \der{S}{x}.
\end{equation}

Taking into account that

\begin{eqnarray}
\label{p-expansion}
&&p = z + \sum_{j=1}^{\infty} \frac{1}{z^{j}} \der{S_{j}}{x}, \\
&&\der{S}{T_{n}} = z^{n} + O\left(\frac{1}{z}\right); ~~~z \rightarrow \infty,
\end{eqnarray}
then

\begin{equation}
\label{omega}
\Omega_{n}(p,T) - \left({\cal Z}^{n}_{0}(p,T)\right)_{+} = {\cal
O}\left(\frac{1}{z} \right); ~~~ z \rightarrow \infty,
\end{equation}

where ${\cal Z}_{0}$ denotes the expansion for $z$ obtained by
inversion of equation~(\ref{p-expansion}), and the symbol $(\cdot)_{+}$
means the polynomial part of the expansion. The left hand sides of
equations~(\ref{omega}) are solutions of $BE$, so
they vanish identically\cite{ragnisco}, i.e.

\begin{equation}
\label{omega*}
\Omega_{n}(p,T) = \left({\cal Z}^{n}_{0}(p,T)\right)_{+}.
\end{equation}

Then, $\Omega_{n}$ can be connected to a suitable
expansion series in terms of the $p$ variable.
Assuming

\begin{equation}
\label{L0expansion}
{\cal Z}_{0}(p,T) = p + \sum_{j=1}^{\infty} \frac{a_{j}(T)}{p^{j}},
\end{equation}
one has

\begin{equation}
\Omega_{1} = p;~~~~\Omega_{2} = p^{2} + 2 a_{1};~~~~\Omega_{3} = p^{3}
+ 3 a_{1} p + 3 a_{2},
\end{equation}

that reproduces equations~(\ref{Stimey}) and (\ref{Stimet}), by identifications

\begin{equation}
u_{0} = 2 a_{1}; ~~~~ V_{0} = 3 a_{2}.
\end{equation}

The set of flows $\Omega_{n}$ represents a set of
quasi-conformal maps for which the dispersionless integrable hierarchy
describes a class of integrable deformations\cite{kon3}. 

\section{Quasi-classical $\bar{\partial}$-dressing method at any order.}
\label{secanyord}
In this section we calculate the corrections at any order for the kernel~(\ref{kernel1ss}).

\begin{equation}
R_{0}(\mu,\bar{\mu};z,\bar{z}) = \frac{i}{2} \sum_{k=0}^{\infty} (-1)^{k}\:
\epsilon^{k-1} \: \delta^{(k)}(\mu-z) \Gamma_{k}(z,\bar{z}). \nonumber
\end{equation}

Let us observe that the independence of $\Gamma_{k,n}$ on the
integration variables $\mu$ and $\bar{\mu}$, does not reduce the
generality of the kernel~(\ref{kernel1ss}). Indeed,
by the delta function 
properties, it provides the same result, up to a redefinition, as for the kernels where
$\Gamma_{k,n}$
depends on integration variables $\mu$ and $\bar{\mu}$.

Substituting the expansion~(\ref{asymptotic0}) into equation~(\ref{DBAR non-local}),  
one obtains, in direct way, the quasiclassical
$\bar{\partial}$-problem at any order. For this purpose, it is
convenient to use the Fa\`a de Bruno
polynomials\cite{deBruno} defined as follows

\begin{equation}
h_{n}[g(x)] = \partial_{x}h_{n-1}[g(x)] + h[g(x)] h_{n-1}[g(x)],
\end{equation}

with $h_{0}[g(x)] = 1$ and $h_{1}[g(x)] = h[g(x)] = \partial_{x}g(x)$,
$\partial_{x} \equiv \der{}{x}$.
Let us note that $h_{n}[g(x)] =
h_{n}[g(x)]\left(\partial_{x}g,\partial^{(2)}_{x}g,\dots,
\partial^{(n)}_{x}g \right) $, in other words, it depends on the derivatives of
$g$ until $n-th$ order.

In our case we have

\begin{equation}
\label{FdB}
h_{l}[\tilde{k} \log S(z,\bar{z})],
\end{equation}
  
for some $\tilde{k} \in \mathbb{N}$. Observing that  

\begin{equation}
h_{l}[\tilde{k} \log S(z,\bar{z})] = \sum_{l'=1}^{l}
C_{l,l'} S^{-l'},
\end{equation}

we denote

\begin{equation}
\label{norm1}
{\cal H}_{s,l} = \frac{C_{s,l}} {l! {\binomial{$\tilde{k}$} {$l$}}},
\end{equation}

where $\tilde{k}$ is an arbitrary integer larger than $l$.
We note that the quantity~(\ref{norm1})
does not depend on $\tilde{k}$.
Now, we introduce the operator
$\hat{W}_{p,q}$ 

\begin{eqnarray}
\hat{W}_{p,q} = W_{p,q}
\partial^{p}_{z}, \nonumber \\
W_{p,q} = \sum_{k=0}^{\infty} (-1)^{k}
\binomial{k}{p}
{\cal H}_{k-p,k-q} \Gamma_{k,n}. \nonumber
\end{eqnarray}

In particular, we have

\begin{eqnarray}
W &=& \hat{W}_{0,0} = \sum_{k=0}^{\infty} (-1)^{k}
{\cal H}_{k,k} \Gamma_{k,n}, \nonumber\\
{\cal H}_{k,k} &=& \left( \der{S}{z} \right)^{k}.\nonumber
\end{eqnarray}

So, we have all ingredients to write the quasiclassical
$\bar{\partial}$-problem at $n-th$ order:
 
\begin{eqnarray}
\label{gen0}
\derbar{S}{z}(z,\bar{z}) &=& W\left(z,\bar{z}, \der{S}{z}\right), \\
\label{gen1}
\derbar{\varphi_{0}}{z} &=& W'
\der{\varphi_{0}}{z} + \frac 1 2 W'' \dersec{S}{z} \varphi_{0},\\
... \nonumber \\
\label{genn}
\derbar{\varphi_{n}}{z} &=& W'
\der{\varphi_{n}}{z} + \frac 1 2 W'' \dersec{S}{z} \varphi_{n}+\sum_{q=1}^{n+1} \hat{W}_{p,q}~\varphi_{n-q+1},
\end{eqnarray}

where

\begin{equation}
W^{(i)} = \frac{\partial^{i}}{\partial \xi^{i}} W\left (z, \bar{z}, \xi \right).  \nonumber  
\end{equation}

It is interesting to observe that equation~(\ref{gen0}) is a
Hamilton-Jacobi type equation, the first order correction is a
homogeneous linear equation, while the next orders are linear equations, but nonhomogeneous ones.
In agreement with the normalization condition~(\ref{boundary}) we have

\begin{eqnarray}
\label{asym2}
\varphi_{0} &\rightarrow& 1+ \frac{\varphi_{0,1}}{z} +
\frac{\varphi_{0,2}}{z^{2}} +
\frac{\varphi_{0,3}}{z^{3}}+\dots , \nonumber\\
\varphi_{1} &\rightarrow& \frac{\varphi_{1,1}}{z} +
\frac{\varphi_{1,2}}{z^{2}} +
\frac{\varphi_{1,3}}{z^{3}}+\dots , \nonumber\\
&\dots& \nonumber \\
\varphi_{n} &\rightarrow& \frac{\varphi_{n,1}}{z} +
\frac{\varphi_{n,2}}{z^{2}} +
\frac{\varphi_{n,3}}{z^{3}}+\dots ; \; n \in \mathbb{N}\setminus \left
\{0\right \}. \nonumber
\end{eqnarray}

The explicit problem at fourth order, i.e. $n=3$ is

\begin{eqnarray}
\label{Seqt}
\derbar{S}{z} &=& W\left(z,\bar{z}, \der{S}{z} \right), \\
\label{transp1}
D_{0}\varphi_{0} &=& 0, \\
\label{transp2}
D_{0}\varphi_{1} &=& D_{1}\varphi_{0}, \\
\label{transp3}
D_{0}\varphi_{2} &=& D_{1}\varphi_{1} + D_{2}\varphi_{0}, \\
\label{transp4}
D_{0}\varphi_{3} &=& D_{1}\varphi_{2} + D_{2}\varphi_{1} + D_{3}\varphi_{0},
\end{eqnarray}   

where

\begin{eqnarray}
\label{D0}
D_{0} &=& \derbar{}{z} - W' \der{}{z} - \frac1 2 W'' \dersec{S}{z}, \nonumber\\
\label{D1}
D_{1} &=& \frac1 2 W'' \dersec{}{z} + \frac1 2 W''' \dersec{S}{z}
\der{}{z} + \frac1 6 W''' \derthree{S}{z} + \frac1 8 W^{(4)}
\left(\dersec{S}{z}\right)^{2},\nonumber\\
\label{D2}
D_{2} &=& \frac1 6 W''' \derthree{}{z} + \frac1 4 W^{(4)}
\dersec{S}{z} \dersec{}{z} + \left(\frac1 6 W^{(4)} \derthree{S}{z} +
\frac1 8 W^{(5)} \left(\dersec{S}{z} \right)^{2} \right) \der{}{z}
+\nonumber \\
&+&\frac{1}{24} W^{(4)} \derfour{S}{z} + \frac{1}{12} W^{(5)} \dersec{S}{z}
\derthree{S}{z} + \frac{1}{48} W^{(6)} \left(\dersec{S}{z}
\right)^{3}, \nonumber\\ 
D_{3} &=& \frac{1}{24} W^{(4)} \derfour{}{z} + \frac{1}{12} W^{(5)}
\dersec{S}{z} \derthree{}{z} + \left(\frac{1}{12} W^{(5)}
\derthree{S}{z} + \frac{1}{16} W^{(6)}
\left(\dersec{S}{z}\right)^{2}\right) \dersec{}{z} + \nonumber \\
&+& \left(\frac{1}{24} W^{(5)} \derfour{S}{z} + \frac{1}{12} W^{(6)}
\dersec{S}{z} \derthree{S}{z} + \frac{1}{48} W^{(7)}
\left(\dersec{S}{z}\right)^{3}  \right) \der{}{z} +\nonumber \\
&+& \frac{1}{120} W^{(5)} \derfive{S}{z} + \frac{1}{72} W^{(6)}
\left(\derthree{S}{z} \right)^{2} + \frac{1}{48} W^{(6)} \dersec{S}{z}
\derfour{S}{z} + \frac{1}{48} W^{(7)}\left(\dersec{S}{z}\right)^{2}
\derthree{S}{z} + \nonumber \\
&+& \frac{1}{384} W^{(8)} \left(\dersec{S}{z} \right)^{4}. \nonumber
\end{eqnarray}


We stress the strong analogy between equations(\ref{Seqt})-(\ref{transp4}) and
the Hamilton-Jacobi and higher transport equations which  arise in the
semi-classical approximation\cite{maslov} and theory of geometric
asymptotics\cite{stern}. Anyway, the main difference with the cited
approaches is that in our case the role of the phase function is
played by the function $S$ that is a complex valued one rather than a
real valued one.




\section{First order contribution.}
\label{secfirst}
The $BE$'s properties are very useful for the study of higher
order corrections too\cite{kon2}. This analysis has been initiated
in the paper\cite{kon2}. Some results of \cite{kon2} are presented here for completeness.
Fixed an arbitrary solution $S(z,\bar{z},T)$ of equation~(\ref{Seqt}),
let $\varphi_{0}$ and $L\varphi_{0}$ be two solutions of
equation~(\ref{transp1}), where $L$ is a suitable linear operator depending
on time variables. The ratio $L\varphi_{0}/\varphi_{0}$ satisfies the
$BE$

\begin{equation}
\derbar{}{z}\left(\frac{L\varphi_{0}}{\varphi_{0}} \right) = W' \der{}{z}\left(\frac{L\varphi_{0}}{\varphi_{0}} \right).
\end{equation}

Then, choosing $L$ such that

\begin{equation}
\label{linear1}
L\varphi_{0} \rightarrow 0; ~~~z \rightarrow \infty
\end{equation} 

as a result of the Vekua's theorem, one gets

\begin{equation}
\label{linear2}
L\varphi_{0} = 0; ~~~ \forall z \in \mathbb{C}.
\end{equation}

In particular, for
the first equation of the KP hierarchy we have

\begin{eqnarray}
\label{phitimey}
L_{1}\varphi_{0} &\equiv& \left(\der{}{y} - 2 \der{S}{x}\der{}{x} -
\dersec{S}{x} - u_{1}(x,y,t) \right) \varphi_{0} = 0, \\
\label{phitimet}
L_{2}\varphi_{0} &\equiv& \left[ \der{}{t} - \left(3 \left(\der{S}{x}
\right)^{2} + \frac3 2 u_{0} \right)\der{}{x} - 3 \der{S}{x}
\dersec{S}{x} - \frac3 2 \der{S}{x} u_{1}+ \right. \nonumber\\
&&~~~~~~~~~~~~~~~~~~~~~~~~~~~~~~~~~~~~~~~~~~\left. - \frac3 4
\der{u_{0}}{x} - V_{1}  \right ] \varphi_{0} = 0,
\end{eqnarray} 

where
\begin{equation}
\label{u1def}
u_{1} = -2 \der{\varphi_{0,1}}{x}; ~~~\der{V_{1}}{x} = \frac3 4
\der{u_{1}}{y}, 
\end{equation}
are extracted from the condition~(\ref{linear1}) and $u_{0}$ is
an arbitrary solution of equation~(\ref{dKP}).

Setting to zero the $1/z$-power expansion's coefficients in~(\ref{phitimey}) and (\ref{phitimet}), we find the
first dispersive correction to dKP equation

\begin{equation}
\label{firstcorr}
\dermixd{u_{1}}{t}{x} = \frac3 2 \dersec{}{x}\left(u_{0} u_{1} \right)
+ \frac3 4 \dersec{u_{1}}{y}.
\end{equation}
 
Let us note that $u_{1}$ satisfies the equation defining the
symmetries of dKP equation~(\ref{dKP}). Indeed, considering equation~(\ref{dKP}) for
$u_{0}+\delta u_{0}$, one, obviously, gets the equation

\begin{equation}
\label{symodd}
\frac{\partial^{2} \left(\delta u \right)}{\partial t \partial x} = {\cal
S}_{0} \left( \delta u \right),
\end{equation}

where

\begin{equation}
{\cal S}_{0}( \cdot ) = \frac3 2 \dersec{}{x} (u_{0}  \cdot)
+ \frac3 4 \dersec{}{y}( \cdot ),
\end{equation}

that coincide with~(\ref{firstcorr}). 

Now we will present some new results concerning the first order corrections.

{\theorem \label{theo1} Let $A_{n}$ and $C_{n}$ be the
differentiable functions defined by

\begin{eqnarray}
A_{n} &=& A_{n}\left(p,T \right) = \left(\dertot{\left({\cal Z}_{0}^n \right)}{p} \right)_{+}, \nonumber\\
C_{n} &=& C_{n}\left(p,\partial_{x}p,T \right) = \frac1 2
\dertot{A_{n}}{x}\left(p,T\right), \nonumber
\end{eqnarray}

where ${\cal Z}_{0}$ is given by~(\ref{L0expansion}), $B_{n} =
B_{n}\left(p,T \right)$ an arbitrary differentiable function, $\varphi_{0}$ some solution of
equation~(\ref{transp1}) and the linear operator $L^{(n)}$ is given by
\begin{equation}
L^{(n)} = \der{}{T_{n}} - A_{n} \der{}{x} - B_{n} - C_{n}.
\end{equation}

Then $L^{(n)} \varphi_{0}$ is also the solution of equation~(\ref{transp1}).}

{\proof By the use of equation~(\ref{Seqt}) and $p =
\der{S}{x}(z,T)$ it's immediate to verify that  

\begin{eqnarray}
\label{An}
\derbar{A_{n}}{z} &=& W' \der{A_{n}}{p} \der{}{z}\der{S}{x} = W' \der{A_{n}}{z}, \\
\label{Bn}
\derbar{B_{n}}{z} &=& W' \der{B_{n}}{p} \der{}{z}\der{S}{x} = W'
\der{B_{n}}{z},\\
\label{Cn}
\derbar{C_{n}}{z} &=& W' \der{C_{n}}{p} \der{}{z}\der{S}{x} + W'
\der{C_{n}}{\left(\partial_{x} p \right)} \dersec{S}{x}+  W''
\der{C_{n}}{\left(\partial_{x}p\right)} \left(\der{}{z}\der{S}{x}
\right)^2. 
\end{eqnarray}
Differentiating equation~(\ref{omega1}) with respect to $z$, we get

\begin{equation}
\label{diffz}
\der{}{z}\left(\der{S}{T_{n}} \right) = A_{n}
\der{}{z}\left(\der{S}{x} \right). 
\end{equation}

Moreover, the definition of the function $C_{n}$

\begin{equation}
\label{Cndef}
C_{n} = \frac1 2 \dertot{A_{n}}{x} = \frac1 2 \der{A_{n}}{p}
\der{p}{x} + \frac1 2 \der{A_{n}}{x},
\end{equation}

implies that

\begin{equation}
\label{Cn2}
\der{C_{n}}{\left(\partial_{x}p \right)} = \frac1 2 \der{A_{n}}{p}.
\end{equation}

Calculating $D_{0}\left(L^{(n)} \varphi_{0} \right)$,
where $D_{0}$ is given in equation~(\ref{D0}), and exploiting
equations~(\ref{Seqt}),(\ref{transp1}) and the set of
equalities~(\ref{An})-(\ref{diffz}),(\ref{Cn2}), one finds 

\begin{equation}
D_{0}\left(L^{(n)} \varphi_{0} \right) = 0.
\end{equation}

This completes the proof. $\Box$ }

Based on the theorem~(\ref{theo1}), we choose, in particular, 

\begin{equation}
B_{n} = \left(\eta(p,T) \dertot{\left({\cal Z}_{0}^n \right)}{p}
\right)_{+} 
\end{equation}

where $\eta$ is a series defined by

\begin{equation}
\label{etaexpansion}
\eta(p,T) = \sum_{j=1}^{\infty} \frac{b_{j}(T)}{p^j}.
\end{equation}

A choice of the coefficients in equation~(\ref{L0expansion}) and
(\ref{etaexpansion}) in such way that $L \varphi_{0} \rightarrow 0$
for $z \rightarrow \infty$, together with the previous arguments on the
{\em BE}, gives us the following set of linear problems in
time variables

\begin{equation}
L^{(n)} \varphi_{0} = 0.
\end{equation} 

These equations can be rearranged by analogy with the leading order
case, in the form

\begin{equation}
\label{flows1}
\der{\log \varphi_{0}}{T_{n}}(z,T) = \Lambda_{n}
\left(q(z,T),p(z,T),\partial_{x}p(z,T),T \right); ~~~z \in \mathbb{C}\setminus {\cal D}, 
\end{equation} 

where

\begin{eqnarray}
\Lambda_{n} &=& \left( {\cal L}_{1} \dertot{\left({\cal Z}_{0}^{n}
\right)}{p} + \frac1 2 \dertot{}{x} \dertot{\left({\cal Z}_{0}^{n}
\right)}{p}  \right)_{+}, \\
{\cal L}_{1} &:=& q + \eta(p,T),\nonumber \\
q &:=& \der{\log \varphi_{0}}{x}. \nonumber
\end{eqnarray}

Taking $n = 1,2,3$

\begin{eqnarray}
\Lambda_{1} &=& q, \nonumber\\
\Lambda_{2} &=& 2 p q + 2 b_{1} + \partial_{x}p,\nonumber \\
\Lambda_{3} &=& 3 p^{2} q + 3 a_{1} q + 3 b_{1} p + 3 p \partial_{x}p +
\frac3 2 \partial_{x}a_{1} + 3 b_{2}, \nonumber
\end{eqnarray}

we reproduce equations~(\ref{phitimey}) and (\ref{phitimet}) and, consequently, the first order dispersive correction to the dKP equation by the
identifications

\begin{eqnarray}
u_{0} &=& 2 a_{1}; ~~~~~ V_{0} = 3 a_{2}; \nonumber \\
u_{1} &=& 2 b_{1}; ~~~~~ V_{1} = 3 b_{2}. \nonumber
\end{eqnarray}


\section{Higher order contributions.}
\label{sechigher}
Unlike the first order, the higher order corrections are
characterized by nonhomogeneous equations. Here we will show how the
$\bar{\partial}$-dressing procedure works in these cases, discussing explicitly the KP
equation. \\
Let us consider a set of functions $\varphi_{0},\varphi_{1},\dots,\varphi_{n}$ satisfying the $\bar{\partial}$
problem at $(n+1)-th$ order and $n+1$ linear operators in time variables
$K^{(0)},K^{(1)},\dots,K^{(n)}$ such that the quantity $\sum_{m=0}^{n}K^{(m)}
\varphi_{n-m}$ satisfies equation~(\ref{transp1}). Then the
ratio

\begin{equation}
\frac{\sum_{m=0}^{n}K^{(m)} \varphi_{n-m}}{\varphi_{0}} 
\end{equation}

is a solution of
$BE$ with complex dilatation $W'$.

Using the same arguments as in the first order case, choosing
$K^{(j)}$, $j = 0,\dots,n$, in such a way that  $\sum_{m=0}^{n}K^{(m)}
\varphi_{n-m} \rightarrow 0$ for $z \rightarrow \infty$, we get the
linear equations

\begin{equation}
\sum_{m=0}^{n}K^{(m)}
\varphi_{n-m} = 0; ~~~~\forall z \in \mathbb{C}.
\end{equation} 

For instance, the second order dispersive corrections to  dKP equation
are associated with the following pair 

\begin{eqnarray}
\label{phi1timey}
K_{1}^{(0)} \varphi_{1} + K_{1}^{(1)} \varphi_{0} &=& 0, \\
\label{phi1timet}
K_{2}^{(0)} \varphi_{1} + K_{2}^{(1)} \varphi_{0} &=& 0,
\end{eqnarray} 

where $K_{1}^{(0)} = L_{1}$ and $K_{2}^{(0)} = L_{2}$ are given
by~(\ref{phitimey}) and (\ref{phitimet}),  

\begin{eqnarray}
K_{1}^{(1)} &=& - \dersec{}{x} - u_{2}, \\
K_{2}^{(1)} &=& - 3 \der{S}{x} \dersec{}{x} - \left(3 \dersec{S}{x} + \frac3 2
u_{1}\right) \der{}{x} - \derthree{S}{x} - \frac3 2 \der{S}{x} u_{2}+
\nonumber \\
&&-\frac3 4 \der{u_{1}}{x} - V_{2},
\end{eqnarray}

and
\begin{equation}
u_{2} = -2 \der{\varphi_{1,1}}{x}; ~~~\der{V_{2}}{x} = \frac3 4
\der{u_{2}}{y},
\end{equation}

are deducted similarly to~(\ref{u1def}).

Equations~(\ref{phi1timey}) and (\ref{phi1timet}) are compatible if and only if
$u_{2}$ satisfies the second order dispersive correction to dKP equation

\begin{equation}
\dermixd{u_{2}}{t}{x} = \frac3 2 \dersec{}{x}\left(u_{0} u_{2}\right)
+ \frac3 4 \dersec{u_{2}}{y} + \frac3 4 \dersec{u_{1}^{2}}{x} + \frac1
4 \derfour{u_{0}}{x}.
\end{equation}

The third order correction corresponds to the compatibility condition for the
pair of the following linear problems

\begin{eqnarray}
\label{pphi1timey}
K_{1}^{(0)} \varphi_{2} + K_{1}^{(1)} \varphi_{1} + K_{1}^{(2)} \varphi_{0}  &=& 0, \\
\label{pphi1timet}
K_{2}^{(0)} \varphi_{2} + K_{2}^{(1)} \varphi_{1} + K_{2}^{(2)} \varphi_{0} &=& 0,
\end{eqnarray} 

where 

\begin{eqnarray}
K_{1}^{(2)} &=& - u_{3}, \\
K_{2}^{(2)} &=& - \derthree{}{x} - \frac{3}{2} u_{3} \der{S}{x} -
\frac{3}{2} u_{2} \der{}{x} - \frac{3}{4} \der{u_{2}}{x} - V_{3},
\end{eqnarray}

with the definitions

\begin{equation}
u_{3} = -2 \der{\varphi_{2,1}}{x}; ~~~\der{V_{3}}{x} = \frac3 4
\der{u_{3}}{y}.
\end{equation}

The equation for $u_{3}$ is of the form  

\begin{equation}
\dermixd{u_{3}}{t}{x} = \frac{3}{2} \dersec{}{x}\left(u_{0}
u_{3}\right) + \frac{3}{2} \dersec{}{x}\left(u_{1} u_{2} \right) +
\frac{3}{4} \dersec{u_{3}}{y} + \frac{1}{4} \derfour{u_{1}}{x}. 
\end{equation}

A simple observation from equations~(\ref{phi1timey}) and (\ref{phi1timet})
allows us to rewrite them in a compact form, suggesting the
generalization to any
order. In order to realize that, we introduce
the following operator

\begin{equation}
\Delta^{k}[S;\theta] = \left\{ \parbox[c]{5cm}{$e^{-\theta S}
\frac{\partial^{k}}{\partial x^{k}} e^{\theta S}$ ~~~~~if~~~ $k \geq
0$  \\ $0$ ~~~~~~~~~~~~~~~~~~if~~~ $k<0$} \right. 
\end{equation}

Noting that for $k \geq 0$

\begin{equation}
\frac{d^{r}}{d \theta^{r}} \Delta^{k}[S;\theta] =
\left[ \dots \left[\frac{\partial^{k}}{\partial x^{k}}, S \right],
\dots, S \right] \;\;\;\;\; r\;brackets
\end{equation}

the transport equations can be written as follows

\begin{eqnarray}
\der{\varphi_{n}}{T_{k}} &=& \sum_{r=0}^{k-1} \frac{1}{r!}
\frac{d^{r}}{d \theta^{r}} \Delta^{k}[S;0] \varphi_{n-k+r+1} +
\frac{k}{2} \sum_{j=0}^{n} \sum_{r=0}^{k-2} u_{j+r} \frac{d^{r}}{d
\theta^{r}} \Delta^{k-2}[S;0] \varphi_{n-j} + \nonumber \\
&+& \sum_{j=0}^{n}
\left(\frac{3}{4} \der{u_{j}}{x} + V_{j+1} \right) \Delta^{k-3}[S;0] \varphi_{n-j},
\end{eqnarray}

for $k = 1,2,3$. \\

It would be nice to get a similar formula for higher times too.

At last, as we noted above, a solution $u_{1}$ of the first dispersive
equation is defined up to a symmetry of the dKP equation.
Using this freedom one can fix a ``gauge'' putting $u_{1} =
0$. Equations for higher corrections imply that the gauge $u_{n} = 0$
for each odd $n$ is an admissible one.
In such a gauge the even
order corrective equations take the form

\begin{equation}
\frac{\partial^{2} \left(u_{n} \right)}{\partial t \partial x} = {\cal
S}_{0} \left(u_{n}\right) + f_{n}, ~~~~n~ even.
\end{equation}

That is a nonhomogeneous equation with the corresponding homogeneous
one given by the symmetry equation. 
By virtue of the Fredholm's theorem, the nonhomogeneous term must be orthogonal
to the solutions $\delta^{*}u$ of adjoint homogeneous equation, i.e.

\begin{equation}
\label{hortogonality}
\int \delta^{*} u(P) f_{n}(P) dP = 0, \;\;\; P = (x,y,t,\dots).
\end{equation} 

One can check that for regular solutions of the KP equation, the
condition~(\ref{hortogonality}) is satisfied. Quite different
situation takes place for the singular sector of the dKP equation. For
instance, for breaking waves solutions for which $\delta u_{0} \equiv
\der{u_{0}}{x} \rightarrow \infty$, the
condition~(\ref{hortogonality}) breaks.
So, there are obstacles to construct the higher order
quasi-classical corrections for dKP equation originated from its own
singular sector.
Such an obstacle is analogous to a typical one appearing in the
construction of 
global asymptotics in  the semi-classical approximation to the quantum
mechanics\cite{maslov} and in the study of the wave corrections to  geometrical
optics\cite{stern}, because of existence of caustics. In the dispersionless KdV
case this kind of 
problem induces a stratification of the affine space of times providing a
classification of the singularities\cite{kodkon}.   
The problem of obstacles for the construction of the higher order
corrections to dKP equation will be considered elsewhere. 

\section{A Whitham type structure.}
\label{secLax}
It is well known that the dispersionless limit of some integrable
hierarchies admits a symplectic structure\cite{krich1}-\cite{taktak2}. 
In particular, one can see that starting with the function  $S(z,T)$
outside the circle ${\cal D}$ it is possible to introduce a 2-form closed
$\omega_{0}$ in terms of the p-variable such that

\begin{equation}
\label{twoform}
\omega_{0} = d\Omega_{n}\left(p,T \right) \wedge dT_{n} = d{\cal
Z}_{0}\left(p,T\right) \wedge d\tilde{{\cal M}}_{0}\left(p,T \right),
\end{equation}

where the sum over index $n$ is assumed. Here the last member
contains the pair of Darboux coordinates ${\cal Z}_{0}$ and
$\tilde{{\cal M}}_{0}\left(p,T \right)={\cal M}_{0}\left({\cal
Z}_{0}(p,T),T \right) = {\cal M}_{0}\left(z,T \right)$ and the function

\begin{equation}
{\cal M}_{0}(z,T) := \der{S}{z}(z,T)
\end{equation}

is usually called Orlov's function\cite{orlov}.

From~(\ref{twoform}), it follows that

\begin{eqnarray}
\label{string}
1 &=&\left \{{\cal Z}_{0}, \tilde{{\cal M}}_{0} \right \}_{p,x}, \\ 
\label{Lax1}
\der{{\cal Z}_{0}}{T_{n}} &=& \left \{\Omega_{n}, {\cal Z}_{0} \right
\}_{p,x},\\
\label{Lax2}
\der{\tilde{{\cal M}}_{0}}{T_{n}} &=& \left \{ \Omega_{n}, \tilde{{\cal M}}_{0} \right \}_{p,x}.
\end{eqnarray}

where the Poisson bracket is defined as follows 

\begin{equation}
\left\{ f(\alpha,\beta), g(\alpha,\beta) \right \}_{\alpha,\beta} =
\der{f}{\alpha} \der{g}{\beta} - \der{g}{\alpha} \der{f}{\beta}. 
\end{equation}

Equation~(\ref{string}) is usually referred to as {\em string
equation}\cite{taktak1} and
equation~(\ref{Lax1}) is the dispersionless Lax pair.
Moreover, equation~(\ref{twoform}) implies the zero curvature condition 

\begin{equation}
\label{square}
\omega_{0} \wedge \omega_{0} = 0.
\end{equation}

That is another way to write the compatibility condition between
equations~(\ref{Lax1}) or~(\ref{Lax2}) ; indeed, expanding the total differentials,
one gets from equation~(\ref{square}) the set of equations

\begin{equation}
\label{whitam0}
\der{\Omega_{n}}{T_{m}} - \der{\Omega_{m}}{T_{n}} +
\poisson{\Omega_{n}}{\Omega_{m}}{p}{x} = 0,
\end{equation}

that are the universal Whitham hierarchy equations\cite{krich1}. In particular, for $n = 2$
and $m = 3$ one obtains equation~(\ref{dKP}).

The construction of the infinite set of flows in equation~(\ref{flows1}),
allows us to describe the first correction in a way analogous to the zero
curvature condition~(\ref{square}) . Let us start with the total differential of $\log
\varphi_{0}$ 

\begin{equation}
\label{total2}
d\log \varphi_{0}(z,T) =
\Lambda_{n}\left(q(z,T),p(z,T),\partial_{x}p(z,T),T \right) dT_{n} +
{\cal M}_{1}(z,T) dz.
\end{equation}

The function

\begin{equation}
{\cal M}_{1}(z,T) = \der{\log \varphi_{0}}{z}(z,T) 
\end{equation}

is an analogous of the Orlov's function. Differentiating
equation~(\ref{total2}), one can introduces the 2-form

\begin{equation}
\label{twoform2}
\omega_{1} := d\Lambda_{n}\left(q(z,T),p(z,T),\partial_{x}p(z,T),T
\right) \wedge dT_{n} = dz \wedge d{\cal M}_{1}(z,T).
\end{equation}

The equation~(\ref{twoform}) allows us to identify a Whitham type structure defined by the equation

\begin{equation}
\omega_{1} \wedge \omega_{1} = 0,
\end{equation}

which can be given explicitly

\begin{eqnarray}
\label{whitham1}
&&\der{\Lambda_{n}}{T_{m}} - \der{\Lambda_{m}}{T_{n}} +
\poisson{\Lambda_{n}}{\Lambda_{m}}{q}{x}+
\poisson{\Lambda_{n}}{\Lambda_{m}}{q}{p} \der{p}{x}+
\poisson{\Lambda_{n}}{\Lambda_{m}}{q}{\partial_{x}p} \dersec{p}{x} = \nonumber \\
&&= G_{n}\left(\Lambda_{m}\right) - G_{m}\left(\Lambda_{n}\right),  
\end{eqnarray}

where $G_{n}$ is defined as

\begin{equation}
G_{n}\left(f \right) = \left(\dertot{\Omega_{n}}{x} \der{}{p} +
\dertotsec{\Omega_{n}}{x} \der{}{\left(\partial_{x}p\right)} \right) f. 
\end{equation}

Let us note that for $n=2$ and $m=3$, equation~(\ref{whitham1}) gives,
of course equation~(\ref{firstcorr}).

\section{Concluding remarks and perspectives.}
\label{secend}
In this paper we demonstrated how to generalize the
quasi-classical $\bar{\partial}$-dressing method at any dispersive order.
In particular, the formula~(\ref{flows1}) suggests that a regular
structure survives at higher orders, allowing to define a potential
encoding all informations about the hierarchy at the first dispersive
order, just like in the purely dispersionless case. Moreover the paper
provides us with several intriguing observations which could be
subjects of future study.
In particular, it would be useful to analyze the connection between the
standard theory of the asymptotics approximation and the
quasi-classical $\bar{\partial}$-dressing approach. This could set a
relationship between the source of the obstacles in the construction
of the corrections, discussed in the section~(\ref{sechigher}), and
the caustics problem. 
As far as concerning possible applications it would be of interest to consider the KdV
limit of the KP 
hierarchy to establish a connection with the Dubrovin-Zhang
theory\cite{dubrovin}.   
Another intriguing matter of study and application is associated with
a possible physical interpretation of the solutions of dispersionless
systems as describing integrable dynamics of                   
interfaces\cite{mineev}. In this context, one should 
investigate a quasiconformal dynamics described by the quasiclassical
$\bar{\partial}$ problem  and the corrections~(\ref{gen0}-\ref{genn}).
One could analyze the singular behavior of interfaces'
evolution, by introducing, for instance, a small ``dispersive'' parameter. 
For such a purpose, a deeper study of the singular
sector order by order will be also necessary.

\end{document}